\documentclass[aps,prl,reprint,amsmath,amssymb,superscriptaddress,showpacs]{revtex4-1}
\usepackage{bm}
\usepackage{graphicx}
\usepackage{color}
\usepackage{dcolumn}
\usepackage{kotex}
\usepackage{float}
\usepackage{hyperref}
\usepackage{amsmath}
\usepackage{siunitx}

\begin{document}

\title{Renormalization of spin-excitations in hexagonal HoMnO$_3$ from magnon-phonon coupling}

\author{Taehun Kim}
\author{Jonathan C. Leiner}
\author{Kisoo Park} 
\author{Joosung Oh}
\author{Hasung Sim}
\affiliation{Center for Correlated Electron Systems, Institute for Basic Science (IBS), Seoul 08826, Republic of Korea}
\affiliation{Department of Physics and Astronomy, Seoul National University, Seoul 08826, Republic of Korea}
\author{Kazuki Iida}
\author{Kazuya Kamazawa}
\affiliation{Neutron Science and Technology Centre, Comprehensive Research Organization for Science and Society (CROSS), Tokai, Ibaraki 319-1106, Japan}
\author{Je-Geun Park}\email{jgpark10@snu.ac.kr}
\affiliation{Center for Correlated Electron Systems, Institute for Basic Science (IBS), Seoul 08826, Republic of Korea}
\affiliation{Department of Physics and Astronomy, Seoul National University, Seoul 08826, Republic of Korea}

\date{\today}

\begin{abstract}
Hexagonal HoMnO$_3$, a two-dimensional Heisenberg antiferromagnet, has been studied via inelastic neutron scattering. A simple Heisenberg model with a single-ion anisotropy describes most features of the spin-wave dispersion curves. However, there is shown to be a renormalization of the magnon energies located at around 11 meV. Since both the magnon-magnon interaction and magnon-phonon coupling can affect the renormalization in a noncollinear magnet, we have accounted for both of these couplings by using a Heisenberg XXZ model with $1/S$ expansions [1] and the Einstein site phonon model [13], respectively. This quantitative analysis leads to the conclusion that the renormalization effect primarily originates from the magnon-phonon coupling, while the spontaneous magnon decay due to the magnon-magnon interaction is suppressed by strong two-ion anisotropy.
\end{abstract}

\maketitle
$Introductions.$ One of the fundamental questions in modern condensed matter physics is to understand how strong correlations among different degrees of freedom affect the otherwise noninteracting energy bands of each individual object. The experimental observation of such effects is key to verifying the predictions of the relevant theoretical frameworks describing the systems in question. Quintessential examples of such effects include the renormalizations and energy shifts in the magnon and phonon spectra, which arise from magnon-magnon and magnon-phonon couplings.

As predicted in theories \cite{PhysRevLett.97.207202,PhysRevLett.96.057201,PhysRevB.74.180403,PhysRevB.88.094407} and subsequently observed in experiments \cite{PhysRevLett.111.257202,PhysRevLett.116.087201}, the spontaneous magnon decay originating from the magnon-magnon interaction has been studied in two-dimensional triangular Heisenberg antiferromagnets (2D THA). Anharmonic terms in the spin Hamiltonian can survive due to the noncollinear magnetic structure of the \ang{120} spin ordered state \cite{Nat.Commun.7.13146}. This leads to some anomalous features such as the strong renormalization of magnon energies and an intrinsic linewidth broadening of magnon spectra due to the finite lifetime of the magnons.

Hexagonal rare-earth manganites $R$MnO$_3$, one of the famous multiferroic materials, are practical candidates for having such couplings. In fact, this material can have both magnon-magnon and magnon-phonon couplings because the spin and lattice degrees of freedom are strongly coupled to each other \cite{Nature.451.805,JKPS.49.5}. Previous inelastic neutron scattering (INS) studies reported that the magnetic excitations of $R$MnO$_3$ with nonmagnetic $R$ ions (Y/Lu) exhibit several of the aforementioned features \cite{PhysRevLett.111.257202,PhysRevLett.99.266604}. These features from the anharmonicity are enhanced by strong magnon-phonon coupling \cite{Nat.Commun.7.13146}. However, the experimental results obtained so far from (Y/Lu)MnO$_3$ cannot be precisely compared with calculations \cite{PhysRevLett.97.207202,PhysRevB.88.094407}. The lowered lattice symmetry due to Mn trimerization \cite{ActaCryst.B72.3} requires a minimum of four exchange interactions for an analysis of the measured spin waves. Given the complexity of the model Hamiltonian, it is practically impossible to carry out nonlinear calculations for (Y,Lu)MnO$_3$ for a quantitative comparison with the experimental data.

\begin{figure}[b]
	\centering
	\includegraphics[width=1.0\columnwidth,clip]{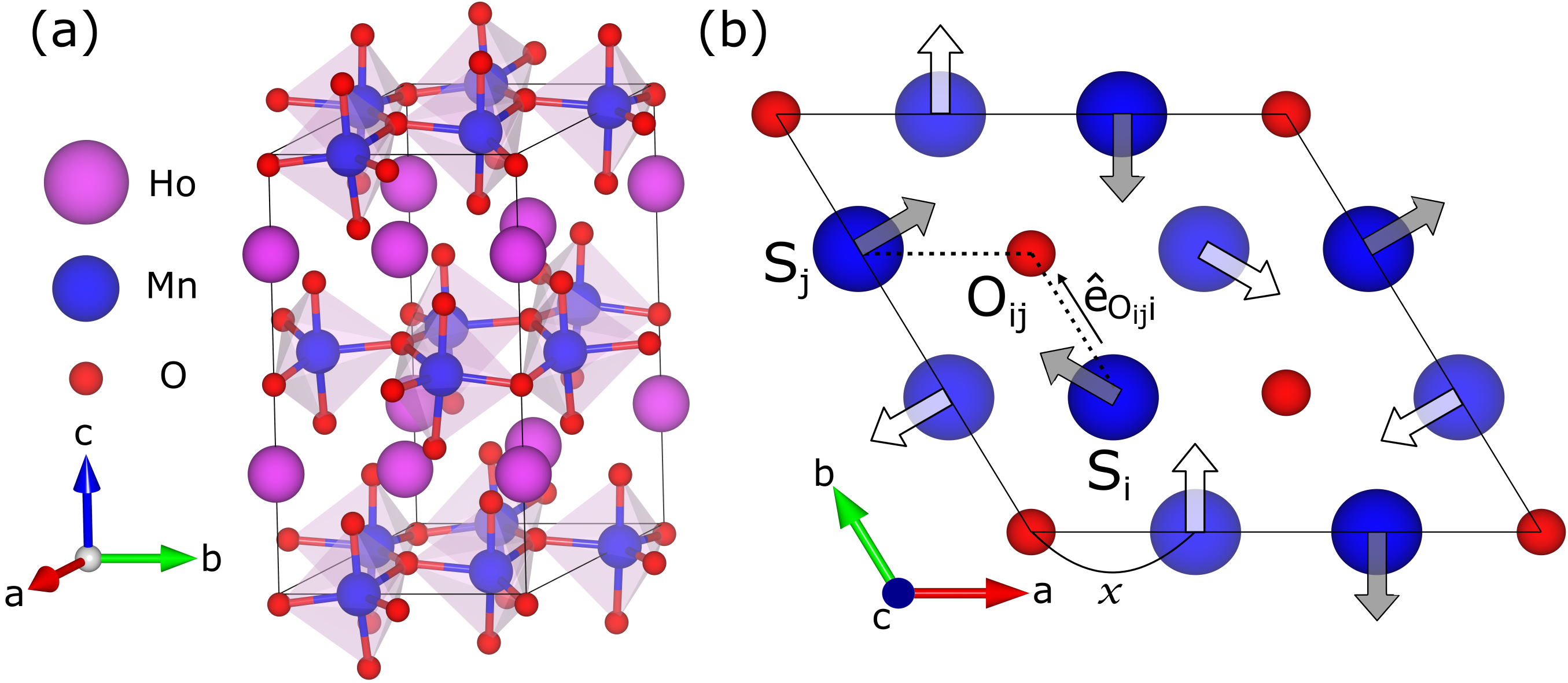}
	\caption{(a) Crystal structure and (b) magnetic structure of HoMnO$_3$. Mn ions with an open arrow are at the $z=0$ plane and those with a solid arrow are at the $z=1/2$ plane. $x$ denotes the distance of the Mn atom from the origin.}
	\label{HMO_structure}
\end{figure}

Hexagonal HoMnO$_3$ as a near perfect 2D THA is an ideal candidate for this purpose, because the Mn position $x$ as shown in Fig. \ref{HMO_structure}(b) is close to 1/3 of the lattice parameter \cite{PhysRevLett.103.067204}. In addition, INS data revealed that the two different exchange couplings distinguished by the Mn position are very similar, within 0.0018 meV, which indicates that the Mn position is close to 1/3. Hence, the 2D frustrated \ang{120} ordered magnetic structure of ideal 2D THA can be manifested with the greatest fidelity in HoMnO$_3$. 

In this Rapid Communication, we have studied the spin dynamics of a hexagonal HoMnO$_3$ single crystal using time-of-flight INS, which allows for the modeling and quantification of both magnon-magnon and magnon-phonon interactions. We compare the data with three different model calculations: first, a simple Heisenberg model within linear spin-wave theory, and then two models that account for higher-order effects from both magnon-magnon \cite{PhysRevLett.97.207202,PhysRevLett.96.057201,PhysRevB.74.180403,PhysRevB.88.094407} and magnon-phonon coupling \cite{PhysRevLett.100.077201}. Using these three models, we are able to accurately model the whole INS spectra of HoMnO$_3$.

$Experimental details.$ A HoMnO$_3$ single crystal was synthesized by using an optical floating zone furnace (Crystal Systems, Japan) with a size of 5$\times$ 5 $\times$ 22 mm$^3$ and a total mass of about 3 g. The INS experiments with this single crystal were carried out using the 4SEASONS time-of-flight spectrometer at J-PARC \cite{J.Phys.Soc.Jpn.80}. The sample was aligned in the $(HHL)$ plane and the incident neutron beam was set as $\vec{k_i} \parallel$ $(00L)$. The frequency of the Fermi chopper was fixed at 250 Hz, which with the multirep mode \cite{J.Phys.Soc.Jpn.78} allows for simultaneous data collection with incident energies of $E_i$ = 60, 30, 18, 12 and 8.5 meV. All spectra were taken at 4 K and the full width at half maxima (FWHM) of the elastic peak obtained by fitting the Lorentzian function were 0.21, 0.31, 0.59, 1.1, and 2.5 meV for $E_i$ = 8.5, 12, 18, 30, and 60 meV, respectively.

As outlined in the Introduction, three different model calculations have been employed to describe the shape and intensity of the spin dispersion curves. First, the simple Heisenberg model including a nearest-neighbor exchange interaction $J$ and a single-ion anisotropy $D$ was used within linear spin-wave theory (LSWT) with the following equations, 
\[
H_{Heisenberg} = J\sum_{<ij>} \vec{S_i}\cdot\vec{S_j}+D\sum_{i} (S_i^z)^2 \tag{1} \label{LSWT}
\]
The SPINW software library was used to calculate the dynamical structure factor with this model \cite{JPCM.27.166002}.

Second, in order to include the effect from magnon-magnon interactions, the Heisenberg XXZ model with $1/S$ expansions \cite{PhysRevLett.97.207202} is considered with the exchange interaction $J$ and two-ion anisotropy $\Delta$ = $J_z$/$J$,
\[
H_{XXZ} = J\sum_{<ij>}\left[S_i^x S_j^x + S_i^y S_j^y + \Delta S_i^z S_j^z\right] \tag{2} \label{XXZ}
\]
This Hamiltonian can also be written by introducing single-ion anisotropy, but here we are considering only the two-ion anisotropy because of the simplicity it offers when considering anharmonic terms in the spin Hamiltonian.

Finally, the third one is the Einstein site phonon (ESP) model \cite{PhysRevB.74.134409,PhysRevLett.100.077201} based on the exchange-striction scheme in order to apply the magnon-phonon coupling in the Hamiltonian,
\[
H_{ESP} = J\left[ \sum_{<ij>} \vec{S_i}\cdot\vec{S_j} - cS^{2}\sum_i \vec{F_i}^2\right] + D\sum_{i} (S_i^z)^2\tag{3}\label{ESP}
\]
$\vec{F_i}$ is the dimensionless force expressed as $$\vec{F_i} = \sum_{j\in n.n.\ of\ i} \hat{e}_{O_{ij}j}(\vec{S_i}\cdot\vec{S_j})/S^2$$ and $\hat{e}_{O_{ij}j}$ is an unit vector from the Mn site $j$ to the O site $O_{ij}$ [see Fig. \ref{HMO_structure}(b)]. The dimensionless spin-phonon coupling constant denoted as $c$ is described as $\alpha^2JS^2/2K$. Here, $K$ is an elastic constant with a unit of energy and $\alpha$ is an exchange-striction coefficient defined as $\alpha = {{d\over{J}}{\partial J\over{\partial r}}}$, where $d$ is a bond length between Mn and O.

\begin{figure}[t]
    \centering
    \includegraphics[width=1.05\columnwidth,clip]{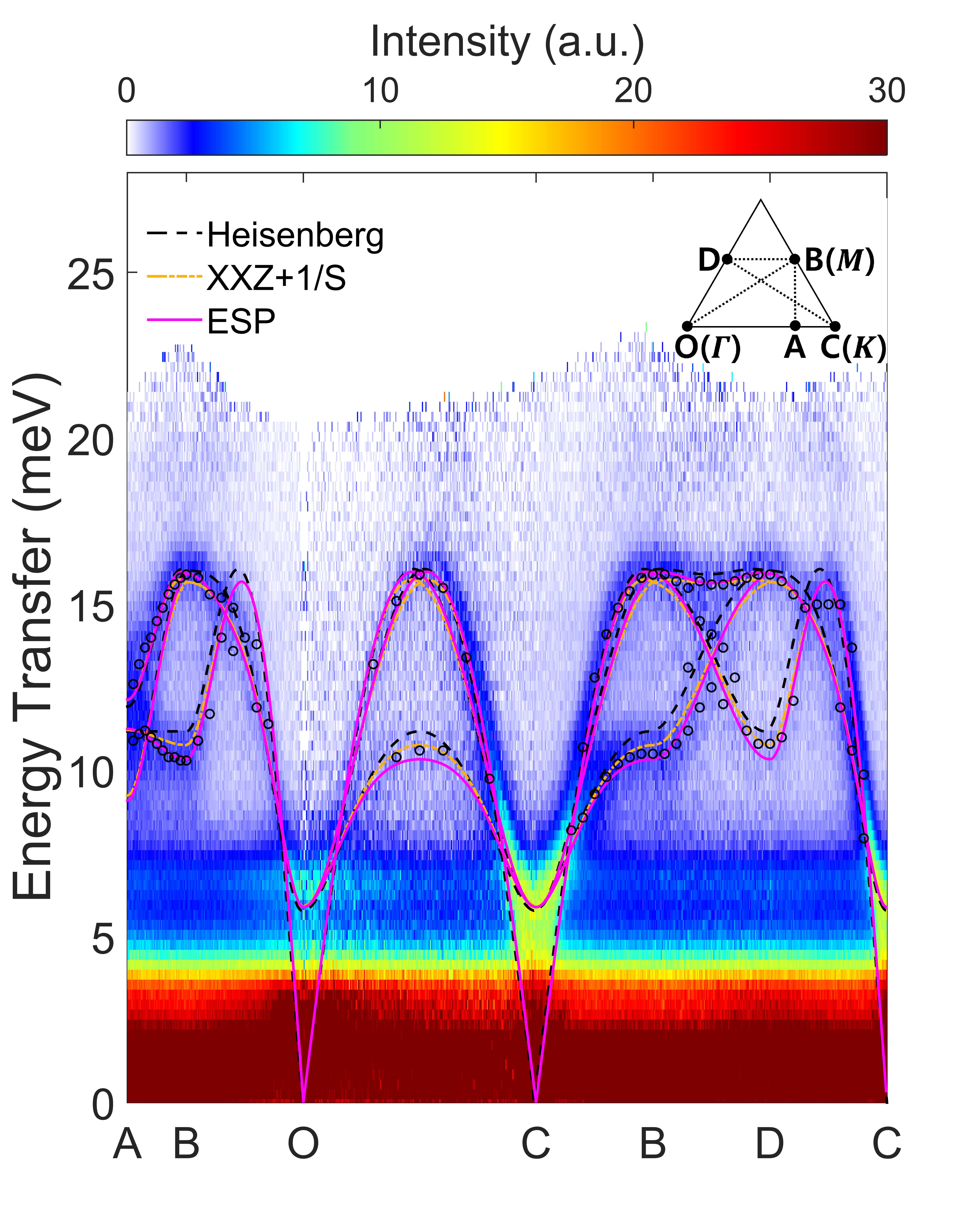}
    \caption{Comparison of the experimental INS spectra of HoMnO$_3$ with calculations using the Heisenberg model (dashed line), the XXZ model with $1/S$ expansions (dashed-dotted line) and the ESP model (solid line), which are described in further detail in the text. The INS data were measured at 4 K and an incident neutron energy of 30 meV. Black circles are fitting positions from constant $Q$ cuts. The inset shows the layout of the momentum position labels.}
    \label{HMO_dispersion}
\end{figure}

\begin{figure}[t]
	\centering
	\includegraphics[width=1.0\columnwidth,clip]{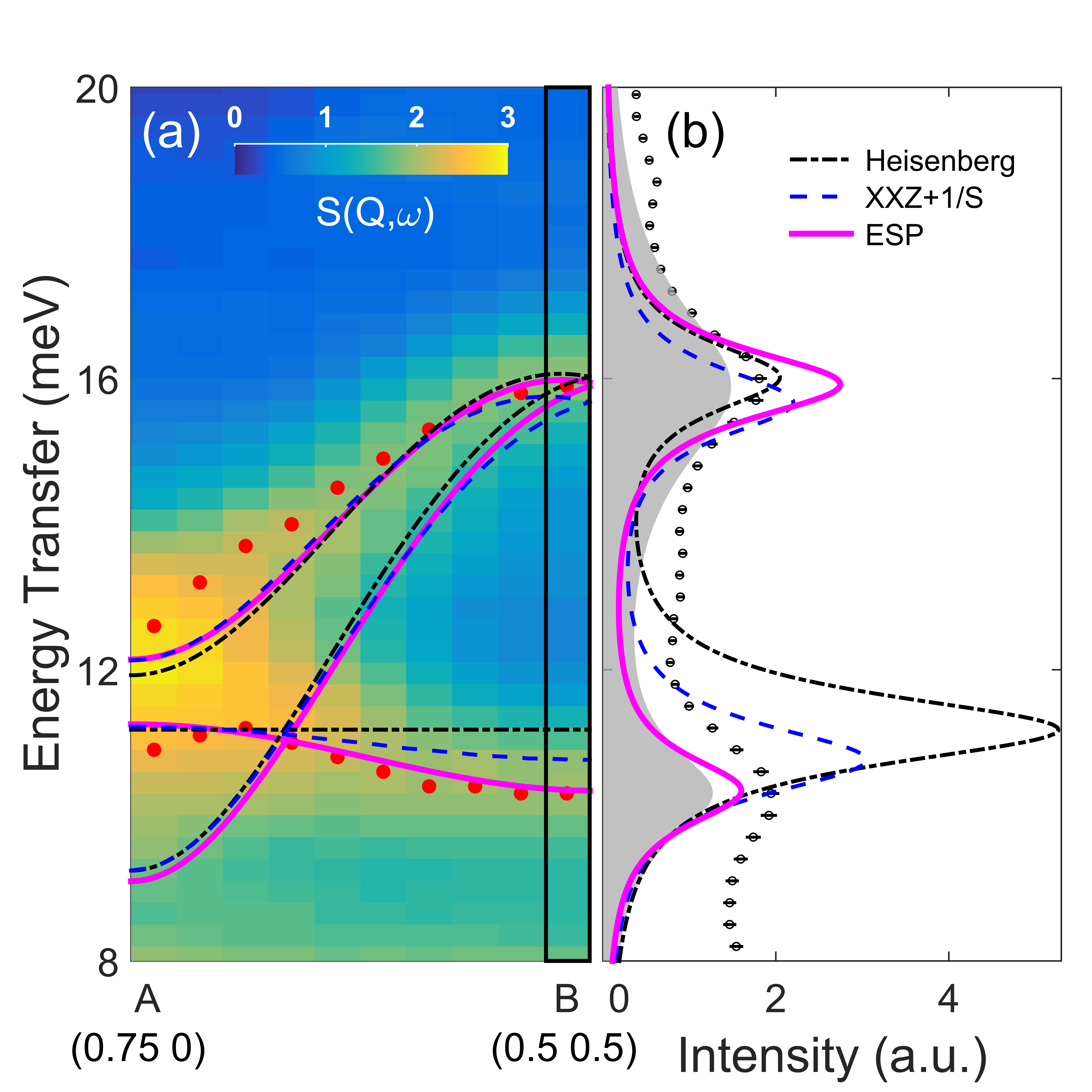}
	\caption{(a) INS data along the $AB$ direction. Red circles indicate the fitted peak center. Three model calculations are also plotted (see the legend). (b) The intensity profile at the $B$ point. All calculations are convoluted with an energy resolution of 0.6 meV with a Lorentzian function. The gray shaded area indicates the two fitted magnon peaks with the background of the data subtracted.}
	\label{HMO_Bpoint}
\end{figure}

$Results and discussion.$ The INS data for HoMnO$_3$ with $E_i$ = 30 meV are summarized in Fig. \ref{HMO_dispersion}. Clear single magnon modes are present in the plot as well as three dispersionless crystal field excitations. The energy transfers of the crystal field excitations from multiple peak fittings are found at 1.7, 3.2, and 6.7 meV, which are consistent with the values reported in Ref. \cite{PhysRevLett.119.227601}. As we expected in an ideal 2D THA, most features of magnons are well captured by the Heisenberg model. Equation \eqref{LSWT} with $J$ = 2.44 meV and $D$ = 0.38 meV is the same as in a previously reported INS study using a triple-axis spectrometer \cite{PhysRevLett.94.087601}, without considering further nearest-neighbor intraplane and interplane exchange interactions. However, there are also some discrepancies that cannot be explained. First, the low-energy magnon dispersion curve located at 11 meV is clearly shifted downward by about 0.8 meV in comparison with the Heisenberg model calculations. In addition, a negative slope in a nominally flat mode was observed along the $AB$ direction, as shown in Fig. \ref{HMO_Bpoint}(a). This feature could not be reproduced by any other type of long-range interaction according to Refs. \cite{PhysRevLett.111.257202,Nat.Commun.7.13146}. We checked that the effect of the exchange anisotropy also could not explain the negative slope within LSWT. The renormalization of magnon energy can be a consequence of both the magnon-magnon interaction and magnon-phonon coupling in a noncollinear magnetic system such as HoMnO$_3$. Therefore, both of these interactions need to be considered in the spin Hamiltonian in order to fully account for each of the anomalous features outlined above.

In order to explain the two aforementioned features, we first used Eq. \eqref{XXZ} with $1/S$ expansions to model the three-magnon interactions. This model clearly reproduces the measured spectrum as well as the renormalization of magnon energy and the downward curve along the $AB$ direction. The best fit parameters for the model are $J$ = 2.7 meV and $\Delta$ = 0.88. In this model, an anharmonic term in the spin Hamiltonian leads to the coupling between the $S^z$ spin component on one sublattice and the $S^{x,y}$ spin components on the other sublattices \cite{PhysRevLett.97.207202,PhysRevLett.111.257202}. The INS data and the calculations together indicate that the spin waves in HoMnO$_3$ also have renormalization expected from the magnon-magnon interactions.

In addition, we also succeeded in reproducing the renormalized magnon dispersion curves by using Eq. \eqref{ESP}. This assumes a coupling between a single magnon and one dispersionless optical phonon branch. As shown in Fig. \ref{HMO_dispersion}, this model also yields a good match with both experimental data and calculated curves from the above XXZ model. The parameters used for the ESP model are $J$ = 2.53 meV and $D$ = 0.38 meV. The obtained dimensionless spin-phonon coupling constant $c$ is 1/12. This value seems to be reasonable, as shown by the fact that the \ang{120} spin ordered ground state can be stabilized up to $c$ = 1/8 in 2D THA \cite{PhysRevLett.100.077201}.

Although both the XXZ model with $1/S$ expansion and the ESP model explain the same magnon dispersion curves, a more significant difference appears in the relative intensities of the dynamical structure factor. As plotted in Fig. \ref{HMO_Bpoint}(b), the ESP model shows the most similar behavior to the observed intensity at the $B$ point as compared with the other two models. More importantly, the relative intensity ratio between the magnon peaks located at 11 and 16 meV is reproduced only by the ESP model. Note that, in order to obtain the best statistics, the data are summed from $L$ = -3 to 3 (taking advantage of the $L$ independence of the data) whereas all calculations are done with $L$ = 0.
\begin{figure}[t]
	\centering
	\includegraphics[width=1.0\columnwidth,clip]{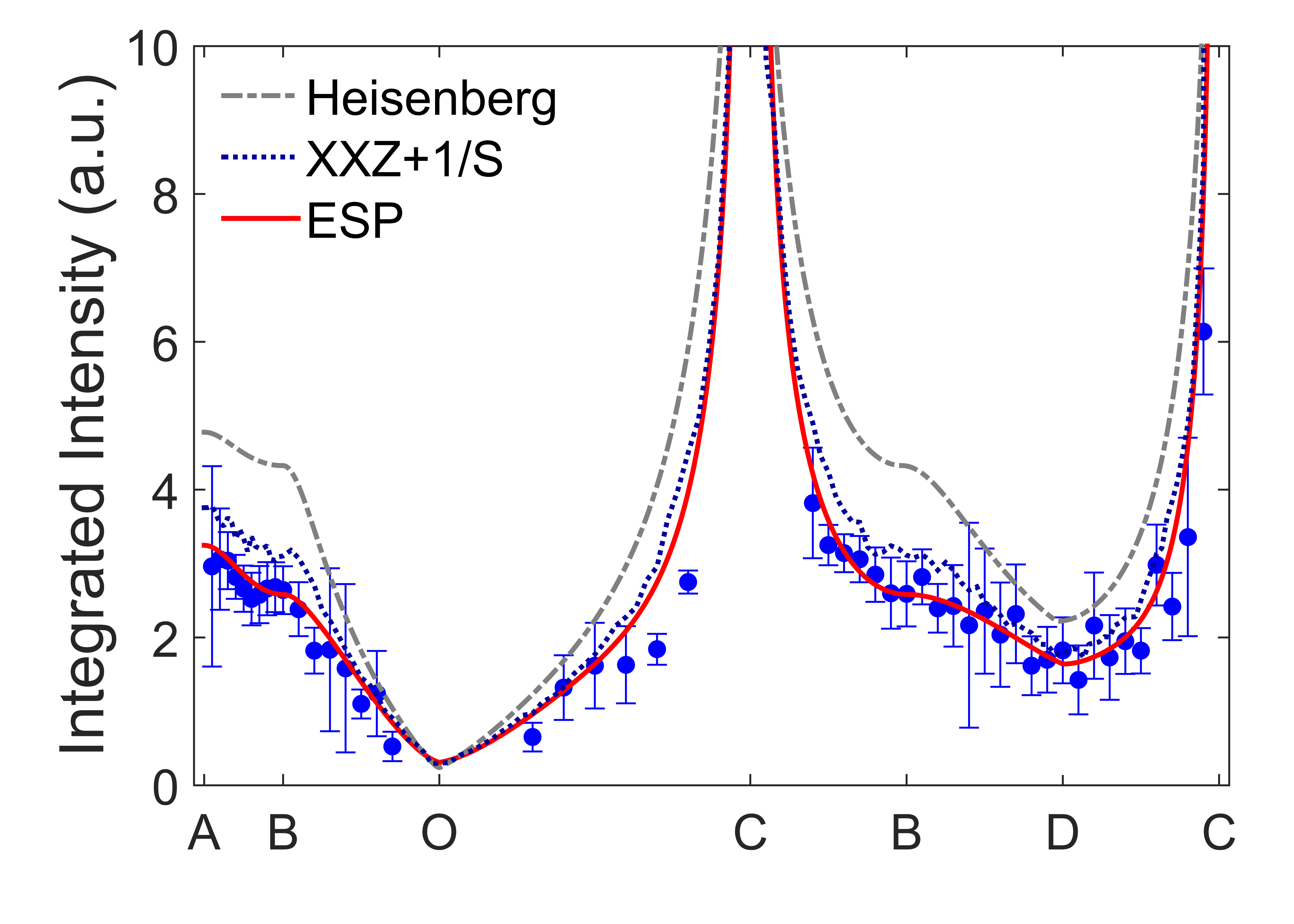}
	\caption{Integrated intensity of the data and calculated dynamical structure factor using three models along the same momentum points as in Fig. \ref{HMO_dispersion}. Solid circles represent the fitted intensity from INS data. Solid, dashed, and dotted lines represent the calculated dynamical structure factor by the ESP model, the Heisenberg model, and XXZ model with $1/S$ expansions, respectively.}
	\label{HMO_int_dispersion}
\end{figure}

As expected from the critical difference on the dynamical structure factor at the $B$ point, the overall intensity dispersion curves are fit remarkably well with the ESP model, as shown in Fig. \ref{HMO_int_dispersion}. We note that all the calculated dynamical structure factors are scaled by a common factor of 0.6. The integrated intensity of the INS data is obtained from the summation of two or three fitted peaks at each $Q$ position. For example, we could fit the magnon peak using two Lorentzian functions at the $B$ point, as shown in Fig. \ref{HMO_Bpoint}(b), and the intensity is obtained by the summation of the two fitted intensities.

Based on this established agreement between experiments and calculations, we suggest that magnon-phonon coupling is the dominant mechanism driving the renormalization away from the Heisenberg model. We note that the spin-phonon coupling constant $c$ we found gives an index of coupling strength in HoMnO$_3$. To obtain a better understanding of how this value for $c$ arises, it is prudent to convert it to the exchange-striction coefficient $\alpha$, where $c$ = $\alpha^2JS^2/2K$ \cite{PhysRevLett.100.077201}.
Therefore, if the elastic constant $K$ is known, a conversion between the two is straightforward. The elastic constant is related to the elastic properties of solids and one of these elastic properties is the elastic stiffness constant. Ultrasonic wave experiments \cite{PhysRevB.76.174426,PhysRevB.83.054418} have revealed the elastic stiffness constants of YMnO$_3$ and HoMnO$_3$, which are on the order of 10$^{11}$ N/m$^2$. This yields an elastic constant on the order of 10 eV. Also, the density functional theory (DFT) calculations for YMnO$_3$ in Ref. \cite{Nat.Commun.7.13146} showed the elastic constant is about 10 eV. Using the same value $K$ = 10 eV, with the assumption that the elastic constants of YMnO$_3$ and HoMnO$_3$ are similar, it follows that the estimated value for $\alpha_{Ho}$ in the case of HoMnO$_3$ is 12.8, which is larger than (Y,Lu)MnO$_3$, $\alpha_Y$ or $\alpha_{Lu}$ = 8 \cite{Nat.Commun.7.13146}.

Moreover, as seen in Fig. \ref{HMO_dispersion}, there seems to be a very diffuse intensity located at around 18 meV. This may be similar to the magneto-elastic excitations observed in (Y,Lu)MnO$_3$ and CuCrO$_2$ \cite{Nat.Commun.7.13146, PhysRevB.94.104421}; (1) the relatively strong intensity near the $B$ and $D$ points and (2) a similar energy transfer (18 meV) above the coherent modes. Such magnetoelastic excitations originate from a hybridization of specific phonon and magnon modes in noncollinear magnets. 

Regarding the magnon-magnon interactions, the one-magnon decay seems to be suppressed in HoMnO$_3$ as compared in LuMnO$_3$. This is actually another clue that supports the magnon-phonon coupling having a major influence on the renormalization. The FWHM of magnon spectra located at 16 meV at the $B$ point is 0.85 meV, which is much smaller than the reported value of 3.5 meV in LuMnO$_3$ \cite{PhysRevLett.111.257202}. Although the FWHM involves an instrumental broadening, that effect is found to be $\sim$1 meV for LuMnO$_3$ and $\sim$0.6 meV for HoMnO$_3$ at each energy transfer. Since the linewidth broadening of the magnon spectra at high energy is directly related to the decay rate from a one-magnon to a two-magnon continuum, such FWHM values also show that the magnon decay is suppressed in HoMnO$_3$.

Another feature in the data related to the suppression of decay is the weak two-magnon continuum signals in the $E_i$ = 60 meV data set. The location at which two-magnon continuum signals are expected to be strongest is around 25 meV at the $B$ point, according to Eq. \eqref{XXZ}. However, no peaklike signals were observed in this location. 
The strongest possible intensity of the two-magnon continuum is calculated to be 3.4\% of single-magnon energy at the $B$ point, which is quite small. This may be explained as the consequence of the strong two-ion anisotropy $\Delta$ = 0.88 in HoMnO$_3$. As pointed out in Ref. \cite{PhysRevLett.97.207202}, the area where the magnon decay is allowed in the first Brillouin zone is completely eliminated at around $\Delta$$\approx$0.92. Therefore, the two-magnon continuum from one-magnon decay is not expected theoretically in HoMnO$_3$, which seems to be consistent with our experimental results. As a result, the HoMnO$_3$ system has strong magnon-phonon coupling and the suppressed decay of one-magnon modes due to the strong two-ion anisotropy. The strong magnon-phonon coupling we found for HoMnO$_3$ may as well be relevant to its magnetoelectric effect as the latter essentially requires a direct coupling between magnons and optical phonons.

$Conclusion.$ We have studied the INS spectra of HoMnO$_3$, the realization of an ideal 2D THA, and compared it with the theoretical calculations of three different models---the Heisenberg model, the XXZ model with $1/S$ expansions, and the ESP model---to quantitatively investigate the effects from magnon-magnon and magnon-phonon interactions. Entire magnon dispersion curves and the features that deviate from the Heisenberg model are well explained by adding such couplings. However, from the calculated dynamical structure factor and the observed suppression of magnon decay, we have concluded that the magnon-phonon coupling effect is dominant in HoMnO$_3$. Quantifying the exchange-striction coefficient $\alpha$, HoMnO$_3$ has a larger value of $\alpha_{Ho}$ = 12.8 than (Y,Lu)MnO$_3$. Noncollinear magnets in principle always exhibit two generic couplings: magnon-magnon and magnon-phonon. In HoMnO$_3$ only the magnon-phonon coupling is highly influential, while the magnon-magnon coupling is strongly suppressed. 



$Acknowledgements.$ We thank C. Broholm and M. Kenzelmann for useful discussions. The inelastic neutron scattering experiment at the J-PARC was performed under user program (Proposal No. 2014B0154). Work at the IBS CCES was supported by the research program of the Institute for Basic Science (IBS-R009-G1).


\end{document}